\pdfoutput=1
\documentclass[11pt]{article}
\usepackage{authblk}
\usepackage{enumitem}
\usepackage{amsmath}
\usepackage{amssymb}
\usepackage{amsfonts}
\usepackage{gensymb}
\usepackage{graphicx}
\usepackage{caption}
\usepackage{hyperref}
\usepackage{float}
\usepackage{braket}
\usepackage{subcaption}
\usepackage{microtype}
\usepackage{ragged2e}
\usepackage{multicol}
\usepackage{lipsum}
\usepackage{subfiles}
\usepackage[utf8]{inputenc}
\DisableLigatures[-]{}
\usepackage[letterpaper,top=1in,bottom=1in,left=1in,right=1in]{geometry}
\usepackage[style=chem-acs,backend=bibtex,articletitle=false]{biblatex}
\graphicspath{ {images/} }
\bibliography{mybib}

\makeatletter
\renewcommand{\paragraph}{%
  \@startsection{paragraph}{4}%
  {\z@}{1ex \@plus 1ex \@minus .2ex}{-1em}%
  {\normalfont\normalsize\itshape}%
}
\makeatother

\title{Pawpyseed: Perturbation-extrapolation band shifting corrections
for point defect calculations}
\author[1]{Kyle Bystrom}
\author[2]{Danny Broberg}
\author[1]{Shyam Dwaraknath}
\author[1,2]{Kristin A. Persson}
\author[2,3]{Mark Asta}
\affil[1]{Lawrence Berkeley National Laboratory, Energy Technologies Area, 1 Cyclotron Road, Berkeley, CA 94720, United States}
\affil[2]{Department of Materials Science and Engineering, University of California, Berkeley, California 94720, United States}
\affil[3]{Lawrence Berkeley National Laboratory, Materials Sciences Division, 1 Cyclotron Road, Berkeley, CA 94720, United States}
\date{\today}

\begin{document}

\maketitle

\begin{abstract}
Significant progress has been made recently in the automation and standardization of \emph{ab initio} point defect calculations. However, the task of developing, implementing, and benchmarking charge corrections for density functional theory (DFT) point defect calculations is still an open challenge. Here we present a high-performance Python package called pawpyseed, which can read PAW DFT wave functions and calculate the overlap between wavefunctions from different structures. Using pawpyseed, we implement a new band shifting correction derived from first order perturbation theory. We benchmark this method by calculating the transition levels of several point defects in silicon and comparing to experimental and hybrid functional results. The new band shifting method can shift single-particle energies to improve transition level predictions and can be automated and parallelized using pawpyseed, suggesting it could be a useful method for high-throughput point defect calculations.
\end{abstract}

\section{Introduction}
There has recently been significant research interest in automation and high-throughput computing for the \emph{ab initio} prediction of point defect properties \cite{pycdt, pylada, coffee}, as well as the development and standardization of methods for accurately predicting point defect properties from semi-local density functional theory (DFT) \cite{Lany2008,Freysoldt2014}. Most first principles studies of point defects use the supercell method: The point defect is embedded in a supercell of the bulk structure, which is repeated infinitely throughout space and modeled with DFT. This method allows point defects to be modeled in robust, high-performance plane-wave DFT codes, but it has several shortcomings. First, the periodic boundary conditions cause unphysical elastic and electronic interactions of the defect with periodic images of itself (finite-size effects). Finite-size effects can be accounted for using electrostatic corrections such as the Freysoldt correction \cite{Freysoldt2009} and Kumagai correction \cite{Kumagai2014}. Second, semi-local DFT functionals drastically underestimate material band gaps. This ``band gap problem" decreases the accuracy of defect calculations by incorrectly describing the location of defect levels in the band gap \cite{Lany2008,Freysoldt2014}. The band gap problem also results in  unrealistic delocalization of electron density \cite{Lany2008,Freysoldt2014}, which causes electrostatic corrections to break down \cite{Komsa2012}. The delocalization issue is generally circumvented using hybrid functional methods \cite{Batista2006,Deak2010,Komsa2011}, which correct the band gap but are too computationally expensive for high-throughput applications. Therefore, a method which can correct band gap-related errors in defect formation energies from semi-local DFT is needed to enable the prediction of defect properties in high throughput.

The current solutions to this problem are a \emph{band filling} correction to account for band dispersion due to charge delocalization and a \emph{band shifting} correction to account for the inaccurate location of defect levels in the band structure. In this work, we present a DFT post-processing software package called pawpyseed that provides tools to develop, automate, and improve band shifting corrections. Specifically, we implement a perturbation theory-motivated band shifting correction, with computationally expensive portions of the code written in C and parallelized for efficiency. To benchmark the performance of this correction against a more basic band shifting technique and other common correction schemes, we use various correction methods to calculate the transition levels of point defects in silicon with important device applications: the single vacancy \cite{Fahey1989} and the boron \cite{Fahey1989}, phosphorus \cite{Fahey1989}, copper \cite{Graff1995,Klose}, and sulfur \cite{Mo2004} substitutionals. The objective is to obtain semi-local DFT transition levels that better approximate those calculated with hybrid DFT or measured in experiments, which would facilitate high-throughput screening studies of point defects.

\subsection{Defect formation energies and band shifting} \label{bandshifting}

Band shifting, the focus of this work, attempts to correct the location of defect levels in the band gap for semi-local DFT calculations \cite{Lany2008}. The general approach is to calculate accurate band edges of the bulk crystal using a higher level of theory, such as hybrid DFT, and then correct the defect energy based on the band edge shifts:
\begin{align}
\Delta E_{CBM}&=E_{CBM,hybrid}-E_{CBM,GGA}
\label{eq:dcbm_eqn}\\
\Delta E_{VBM}&=E_{VBM,hybrid}-E_{VBM,GGA}
\label{eq:dvbm_eqn}
\end{align}

Here, VBM is the valence band maximum, and CBM is the conduction band minimum. The most common use of these band edge shifts is Fermi level renormalization. In the supercell method, the Fermi level is referenced to the VBM, so the formation energy of each defect of charge state $q$ is shifted by $q\Delta E_{VBM}$ as a Fermi level correction. This shifts each transition level by $\Delta E_{VBM}$ relative to the VBM \cite{Persson2005}. This correction neglects that defect levels derived primarily from conduction or valence bands might shift in energy with those bands when the gap is opened, thereby shifting the transition levels by more or less than $\Delta E_{VBM}$ \cite{Lany2008}. An additional level of complexity is to assume that perturbed host states (PHS, shallow levels derived from delocalized bulk valence or conduction bands) shift in energy with the bands they are derived from when the band edges are shifted to the hybrid level of theory \cite{Persson2005,Lany2008}. Following this reasoning, two terms can be added to the defect formation energy. The first is $\Delta E_{CBM}$ times the number of electrons in conduction band PHSs, and the second is $-\Delta E_{VBM}$ times the number of holes in valence band PHSs.

Without additional data, simple PHS shifting is not compatible with high-throughput approaches because it requires identifying perturbed host states. Due to the underestimation of the band gap in GGA, a deep level state derived from the localized atomic orbitals of an impurity might be located near/above the CBM or near/below the VBM in GGA. An automated script would identify such a state as a PHS, but it has been argued that the energy level of a well-localized state on a defect is well-described by semi-local DFT and does not shift with the band edges \cite{Alkauskas2011}. In addition, a state might have some delocalized host band character and some localized atomic state character, which might cause the energy of the state to shift fractionally with the band edges \cite{Alkauskas2011,Lany2008}. These shortcomings motivate a more sophisticated band shifting correction, which is discussed next.

\subsection{The perturbation-extrapolation band shifting correction} \label{pertbandshifting}

In order to extrapolate the defect level from the semi-local DFT band gap to the experimental or hybrid DFT band gap, several methods have been developed that predict how a defect level changes as the band gap changes \cite{Zhang2001}. Most of these methods are based on DFT calculations with a perturbed and unperturbed system, such as an LDA and LDA+U functional \cite{Janotti2005}. However, Bogusławski et al. \cite{Bernholc1995} projected defect levels onto valence and conduction bands and then shifted the levels a percentage of $\Delta E_{CBM}$ based on the percentage conduction character. Using the formalism for a perturbation-extrapolation band gap correction developed by Lany and Zunger \cite{Lany2008}, we introduce a band shifting correction based on the strategy of Bogusławski et al. This band shifting correction adjusts the Kohn-Sham single-particle states, which are the eigenfunctions of the Kohn-Sham Hamiltonian \cite{Kohn1965}. These states are not unique and are only physically significant because they generally give good approximations to single-particle ionization energies, which allows them to be treated as orbitals occupied by a single electron (or electron pair) in an effective potential. Relating shifts in these single-particle energies to formation energies and transition levels is therefore an approximation.

Because the eigenfunctions of a Hamiltonian form a complete basis, a defect wavefunction can be expanded in the bulk wavefunctions:
\begin{equation}
\psi_D(\mathbf{r}) = \sum_{n,\mathbf{k}} \braket{\psi_{n,\mathbf{k}} | \psi_D}
\psi_{n,\mathbf{k}}(\mathbf{r})
= \sum_{n,\mathbf{k}} A_{n,\mathbf{k}} \psi_{n,\mathbf{k}}(\mathbf{r})
\label{eq:def_as_bulksum}
\end{equation}
$\psi_D$ is a defect wavefunction, and $\psi_{n,\mathbf{k}}$ are evaluated in the pristine bulk. Physically, $\psi_D$ is a single state, as opposed to a band of states in k-space, because the defect is not periodic and therefore does not have dispersion. However, it is evaluated at different k-points in the supercell method, so the supercell method defect levels $\psi_{D,\mathbf{k}}$ will be used to ground the band shifting correction in a practical computational framework. Since all wavefunctions at different k-points are orthogonal,
\begin{equation}
\psi_{D,\mathbf{k}}(\mathbf{r}) = \sum_n A_{n,\mathbf{k}} \psi_{n,\mathbf{k}}(\mathbf{r})
\label{eq:def_as_bulksum_sup}
\end{equation}
If the single-particle energy level of $\psi_{D,\mathbf{k}}(\mathbf{r})$ in DFT is $e^0_{D,\mathbf{k}}$, first-order perturbation theory can be used as suggested by Lany and Zunger to calculate a corrected energy \cite{Lany2008}.
\begin{equation}
e_{D,\mathbf{k}} = e^0_{D,\mathbf{k}}
+ \bra{\psi_{D,\mathbf{k}}(\mathbf{r})} \Delta H \ket{\psi_{D,\mathbf{k}}(\mathbf{r})}
\label{eq:pt}
\end{equation}
Here, $\Delta H$ is a correction term that extrapolates from the DFT picture to the true quasiparticle energies. Assuming the diagonal elements of $\Delta H$ in the basis of the bulk wavefunctions are small, Equation \ref{eq:pt} can be expanded
\cite{Lany2008}:
\begin{equation}
e_{D,\mathbf{k}} = e^0_{D,\mathbf{k}} + \sum_n |A_{n,\mathbf{k}}|^2
\bra{\psi_{n,\mathbf{k}}(\mathbf{r})} \Delta H \ket{\psi_{n,\mathbf{k}}(\mathbf{r})}
\label{eq:pt2}
\end{equation}

A rough but potentially useful approximation to $\Delta H$ is a ``band shifting" operator that shifts the energy of a bulk conduction band by $\Delta E_{CBM}$ (Equation \ref{eq:dcbm_eqn}) and the energy of a bulk valence band by $\Delta E_{VBM}$ (Equation \ref{eq:dvbm_eqn}). In fact, not all conduction bands shift by $\Delta E_{CBM}$ and valence bands by $\Delta E_{VBM}$, but this approximation is useful because the defect states of interest are in or near the band gap and most likely composed of states near the band gap. This band shifting operator can therefore shift defect levels calculated in GGA toward the more accurate hybrid DFT energy levels based on how they project onto host bands.

It is useful to define the ``proportion valence" $v_{D,\mathbf{k}}$ and ``proportion conduction" $c_{D,\mathbf{k}}$ as:
\begin{align}
v_{D,\mathbf{k}} &\equiv \sum_{n \in VB}
|\braket{\psi_{D,\mathbf{k}}|\psi_{n,\mathbf{k}}}|^2 \\
c_{D,\mathbf{k}} &\equiv \sum_{n \in CB}
|\braket{\psi_{D,\mathbf{k}}|\psi_{n,\mathbf{k}}}|^2
\end{align}
Here, VB is the set of valence bands and CB is the set of conduction bands. Using these definitions, Equation \ref{eq:pt2} can be expressed simply:
\begin{align}
e_{D,\mathbf{k}} =& e_{D,\mathbf{k}}^0
+ c_{D,\mathbf{k}}\Delta E_{CBM}
+ v_{D,\mathbf{k}}\Delta E_{VBM}
\label{eq:pt_corr}
\end{align}

Since defect levels should not have k-point dispersion, one can also average $v_{D,\mathbf{k}}$ and $c_{D,\mathbf{k}}$ over k-points, which we choose to do in this work. While this type of correction is of interest to the research community based on its presence in recent reviews \cite{Lany2008,Freysoldt2014}, it does not have a standard formalism or open-source implementation. This is partly because the accurate evaluation of the overlap terms $\braket{\psi_{D,\mathbf{k}}|\psi_{n,\mathbf{k}}}$ is computationally difficult in plane-wave DFT with the projector-augmented wave (PAW) method, which is the most commonly used basis set for modern periodic solid calculations. Pawpyseed implements these overlap evaluations as described in Section \ref{overlap_operators}. It also implements a band shifting correction based on Equation \ref{eq:pt_corr}, called the ``Projection shift," which is described in Section \ref{projshift}.

\section{Results and Discussion} \label{results}

\begin{table}[t]
\caption{Errors of predicted transition levels in eV for the DFT calculations performed with $\Gamma$-centered k-point meshes. Format: ``level in eV (error as percentage of band gap)." The band gap for the Fr method is 0.61 eV, while for the other methods it is 1.19 eV. The HSE06 band gap in other studies is 1.16-1.19 eV.
Fr = Freysoldt finite-size correction; FLR = Fermi Level Renormalization; PrS = Projection Shifting (Equations \ref{eq:projshift} and \ref{eq:singleshift1}); DPrS = Delocalized-State Projection Shifting (Equations \ref{eq:projshift} and \ref{eq:singleshift2}).}
\begin{tabular}{|c|c|c|c|c|c|c|}
\hline
Transition Level & Reference & Fr & Fr+FLR & Fr+FLR+PrS & Fr+FLR+DPrS\\
\hline
Vac ++/+	& 0.14 \cite{Spiewak2013}	& -0.08 (-25\%)	& 0.28 (12\%)	& -0.00 (-12\%)	& 0.14 (0\%) \\
Vac +/0		& 0.02 \cite{Spiewak2013}	& -0.00 (-2\%)	& 0.36 (28\%)	& -0.02 (-3\%)	& -0.01 (-2\%) \\
P +/0		& 1.07 \cite{Jagannath1981}	& 0.49 (-15\%)	& 0.85 (-24\%)	& 1.07 (-6\%)	& 1.04 (-8\%) \\
B 0/-		& 0.04 \cite{Fischer1983}	& 0.12 (16\%)	& 0.48 (37\%)	& 0.13 (7\%)	& 0.15 (8\%) \\
Cu +/0		& 0.20 \cite{Sharan2017}	& 0.08 (-4\%)	& 0.44 (20\%)	& 0.16 (-4\%)	& 0.35 (12\%) \\
Cu 0/-		& 0.54 \cite{Sharan2017}	& 0.32 (7\%)	& 0.68 (11\%)	& 0.41 (-12\%)	& 0.58 (3\%) \\
Cu -/--		& 0.97 \cite{Sharan2017}	& 0.63 (20\%)	& 0.99 (0\%)	& 0.77 (-19\%)	& 0.90 (-8\%) \\
S ++/+		& 0.58 \cite{Deak2010}		& 0.10 (-33\%)	& 0.46 (-11\%)	& 0.62 (3\%)	& 0.53 (-5\%) \\
S +/0		& 0.86 \cite{Deak2010}		& 0.43 (-4\%)	& 0.79 (-7\%)	& 0.97 (8\%)	& 0.89 (1\%) \\
\hline
\end{tabular}\\
\label{tab:tlcomp}
\end{table}

To benchmark the perturbative band shifting correction, we used PyCDT \cite{pycdt} to generate structures and charge states for several point defects in silicon. We calculated the DFT total energy of each defect as described in Section \ref{dftcalc} and used PyCDT, pymatgen \cite{pymatgen}, and pawpyseed post-processing tools to calculate the formation energies using four different correction schemes (Table \ref{tab:tlcomp}). The first is the Freysoldt finite-size correction only, the second adds Fermi level renormalization, and the third and fourth use two variants of perturbative band shifting discussed in Section \ref{projshift}. Projection Shifting shifts all defect levels according to Equation \ref{eq:pt_corr}, while Delocalized-State Projection Shifting shifts these levels a fraction of that value based on how delocalized the band is. The justification for the latter approach is that localized defect levels are well-characterized by atomic or molecular orbitals and therefore should not shift in energy with delocalized bulk host states \cite{Alkauskas2011}. Band filling corrections are included for all calculations, although they have a small effect and do not generally improve the transition level predictions. Correction schemes excluding band filling corrections are provided in the SI.

The thermodynamic transition levels were determined from the difference between the formation energies of different charge states. Analysis of formation energies was not done to avoid the complications of chemical potentials. The existing literature on these transition levels contains hybrid DFT (HSE06 \cite{Krukau2006}) and experimental data \cite{Watkins1980,pensl1986,Knack1999,Yarykin2013} in good agreement. The exceptions are the phosphorus and boron substitutionals, for which we could not find hybrid DFT calculations, likely because these defect levels are too shallow to contain in a reasonable-size supercell.

\begin{figure}[t]
\includegraphics[width=0.33\textwidth]{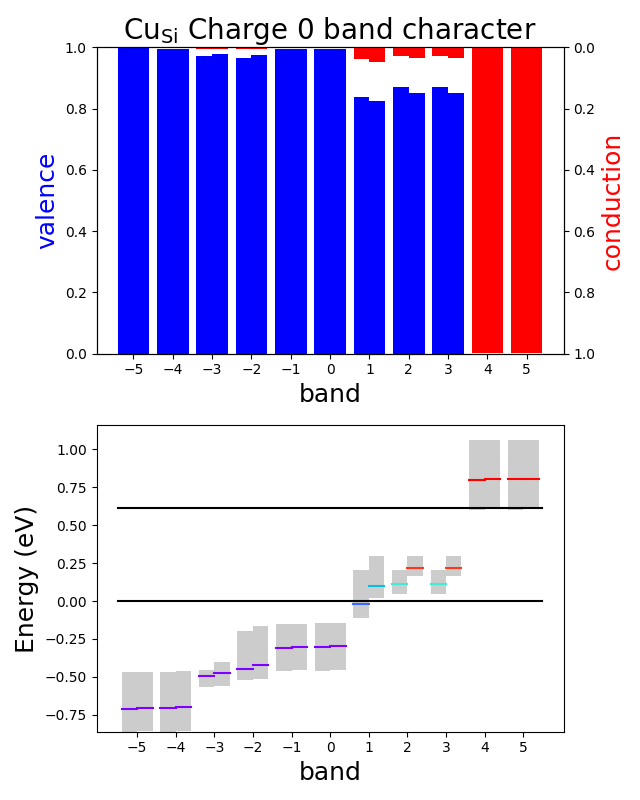}
\includegraphics[width=0.33\textwidth]{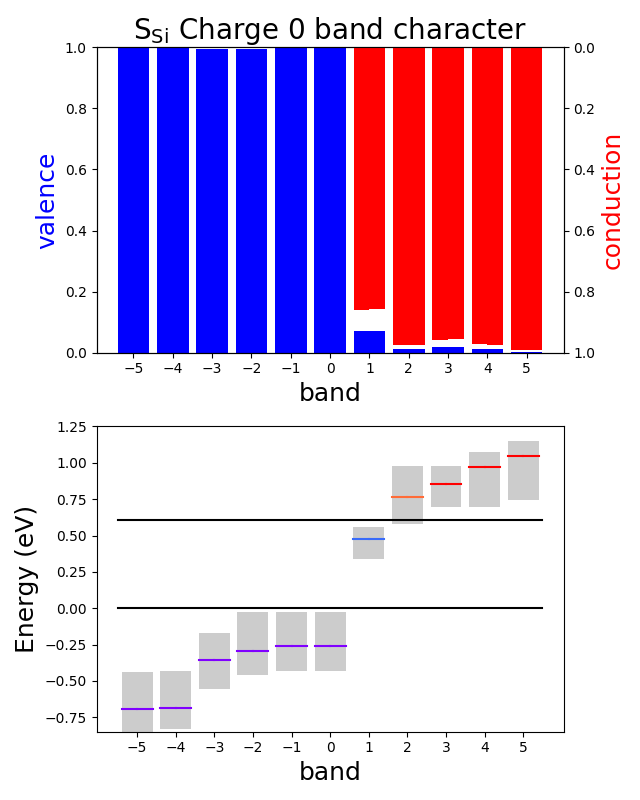}
\includegraphics[width=0.33\textwidth]{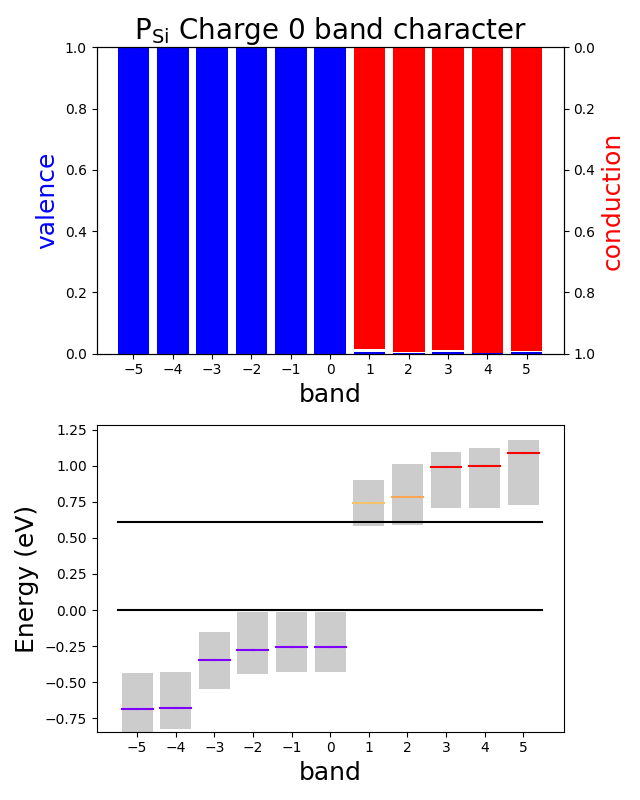}
\caption{$v_D$ (valence band character) and $c_D$ (conduction band character) for states near the band gap of several neutral defects. The upper portion of the plot shows $v_D$ (blue) and $c_D$ (red). The lower part shows the average energy of each spin-polarized band, with the grey bars representing the range of energies of that band over the k-point mesh. The color of the energy levels corresponds to the occupation, with the ``rainbow" color scale provided in Matplotlib \cite{matplotlib} used to clearly illustrate partial occupancies. Violet levels have occupancy 1, and red levels have occupancy 0. The GGA band gap of 0.61 eV is denoted with the black lines. Note that because the bulk basis set is finite and therefore incomplete, $v_D+c_D$ can be less than 1. The missing character, represented by white space between the red and blue bars, is due to the difference between the space spanned by the bulk eigenstates and defect eigenstates. For example, the bulk silicon basis contains no $d$ states, so the white space in the Cu substitutional projection diagram might be due to $d$ state character (or to the Cu $s$ and $p$ atomic orbitals having different shapes than those of Si).}
\label{fig:levels}
\end{figure}

There can be a significant difference between the activation energies measured by deep-level transient spectroscopy (DLTS) and 0 K ionization energies \cite{Wickramaratne2018}. Therefore, we choose to calculate transition level errors relative to the hybrid DFT data except for in the cases of phosphorus and boron, for which we compare to experimental excitation spectra. Because of the good agreement between theory and experiment for these defects, this decision does not significantly impact the interpretation of results. To aid in understanding the action of the Projection Shift on the defect levels, Figure \ref{fig:levels} shows the valence band character $v_D$ and conduction band character $c_D$ for levels near the band gap of the neutral copper, sulfur, and phosphorus substitutionals.

\begin{figure}
\includegraphics[width=0.33\textwidth]{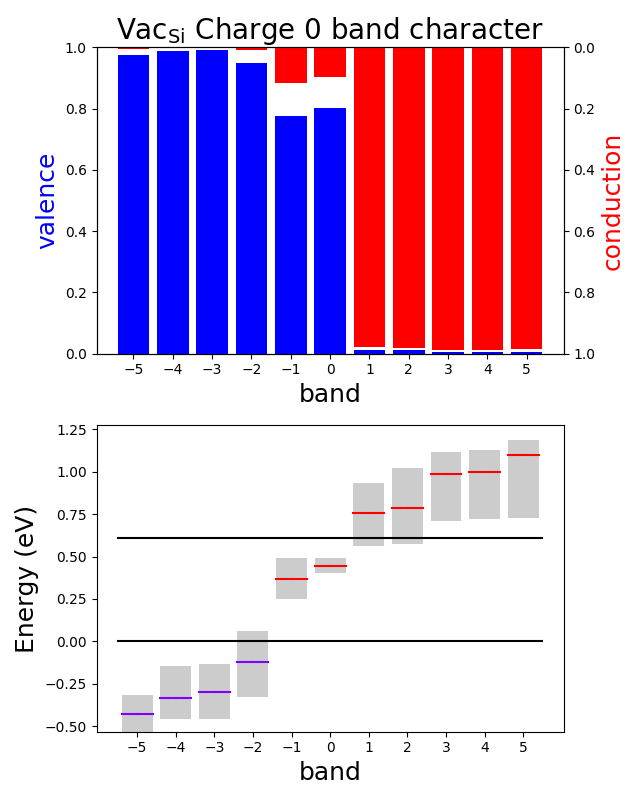}
\includegraphics[width=0.33\textwidth]{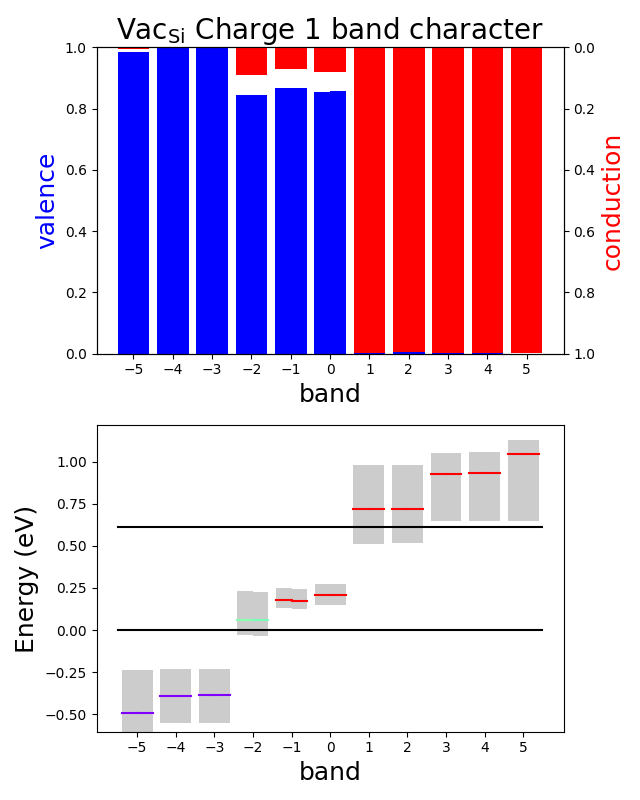}
\includegraphics[width=0.33\textwidth]{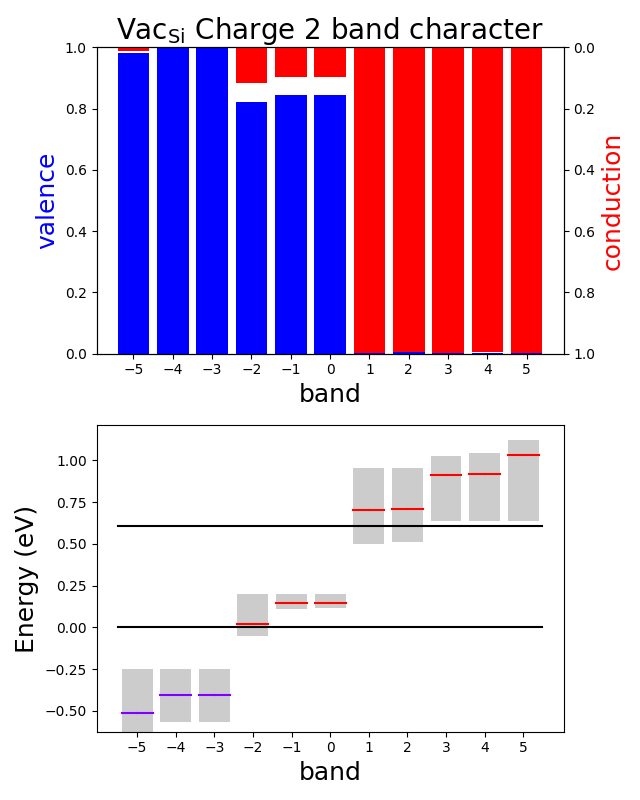}
\caption{$v_D$ and $c_D$ for states near the band gap of several charge states of the silicon vacancy. The highest occupied band of the neutral vacancy has a larger $v_D$ than the same band in the positive charge states, and this band is lower in energy than in the positive charge states.}
\label{fig:vaclevels}
\end{figure}

\subsection{Comparison of non-hydrogenic transition levels to hybrid DFT and experiment}

The single vacancy is a double donor with HSE06 transition levels at 0.14 eV (++/+) and 0.08 eV (+/0) \cite{Spiewak2013}. The ++/+ transition is higher than the +/0 transition due to well-documented and experimentally observed negative-U behavior \cite{Watkins1984}, with experimental levels at 0.13 eV (++/+) and 0.05 eV (+/0) \cite{Watkins1980}. The GGA functional has previously been used to capture the negative-U behavior of the silicon vacancy \cite{Wright2006} without finite-size corrections, but the Freysoldt correction decreases the ++/+ transition level by 0.2 eV, resulting in an incorrect picture of the defect. Interestingly, adding the Projection Shift does result in negative-U behavior for the single vacancy, but the ++/+ level is too low. As shown in Figure \ref{fig:vaclevels}, the occupied defect level in the neutral charge state has more valence band character than in the +1 state (with $v_C=0.95$ for neutral and $v_C=0.82$ for +1). Therefore, the energy of that level shifts lower for the neutral defect and recovers the negative-U character. Because the defect levels in the vacancy are partly localized around the defect, Delocalized-State Projection Shifting shifts the levels less than Projection Shifting. This raises the ++/+ transition to the hybrid DFT value (Table \ref{tab:tlcomp}).

Substitutional copper has three transition levels, which were calculated with the HSE06 functional to be at 0.20 eV (+/0), 0.54 eV (0/-), and 0.97 eV (-/--) \cite{Sharan2017}. Fermi level renormalization overestimates the +/0 and 0/- transition levels significantly. Projection shifting fixes the +/0 level but over-corrects Fermi level renormalization  for the 0/- level compared to HSE06. However, the experimental 0/- level is 0.43 eV \cite{Yarykin2013}, which is 0.11 eV below the HSE06 level and close to the Projection Shift prediction. The -/-- level is predicted accurately by Fermi level renormalization but then shifted -0.22 eV by Projection Shifting (Table \ref{tab:tlcomp}). This suggests that the copper defect levels are too localized to be completely shifted with the host bands. Delocalized-state Projection Shifting partly remedies this issue, but the +/0 level is overestimated and the -/-- level is underestimated.

The sulfur substitutional is a double donor with DLTS levels at 0.546 eV (++/+) and 0.325 eV (+/0) below the CBM \cite{pensl1986}. HSE06 calculations placed the levels 0.54 and 0.26 eV below the CBM, respectively \cite{Deak2010}. The high conduction character of the sulfur defect level (Figure \ref{fig:levels}) matches the expectation from a simple hydrogenic model. Due to this conduction character, the Projection Shift method raises the sulfur transition levels compared to Fermi level renormalization. The Delocalized-State Projection Shifting result is between the other two because the defect level is partially localized around the sulfur.

\subsection{Shallow hydrogenic levels: phosphorus and boron}
The shallowness of the phosphorus and boron substitutionals can decrease the accuracy of supercell calculations due to charge delocalization. These defects are still included because they provide a simple benchmark of whether band shifting methods effectively shift shallow levels with the band edges. The phosphorus donor level is 46 meV below the CBM \cite{Jagannath1981}, and the boron acceptor level is 44 meV above the VBM \cite{Fischer1983}.

As shown in Table \ref{tab:tlcomp}, Fermi level renormalization does not accurately treat shallow defect levels because it does not shift the levels with the valence band edge (for boron/shallow acceptor) or conduction band edge (for phosphorus/shallow donor). The phosphorus and boron levels are at 0.86 eV and 0.44 eV, respectively. Projection Shifting correctly predicts both the boron and phosphorus transition levels because the boron level has nearly 100\% valence character and the phosphorus level has nearly 100\% conduction character (Figure \ref{fig:levels}), so the levels shift with the valence and conduction bands, respectively. The phosphorus and boron levels are highly delocalized, so most of the Projection Shift is preserved by the Delocalized-state Projection Shift.

\subsection{Summary and Conclusions}
We have developed a perturbation-extrapolation correction method for point defect calculations based on projecting defect levels onto the bulk sates. As opposed to previous correction schemes, this new Projection Shift method successfully handles edge cases involving shallow transition levels and single-particle levels. We have published an open-source, parallel Python/C package called pawpyseed that performs the computationally expensive components of the Projection Shift, allowing the method to be further developed and tested. One possible future project is to develop a more robust method for assessing what fraction of the Equation \ref{eq:pt_corr} shift to perform based on how localized a state is. Another is to use Equation \ref{eq:pt_corr} to shift transition levels themselves rather than the single-particle energies in the DFT calculations.

The primary shortcoming of Fermi level renormalization is that the missing physics in semi-local DFT compared to real physical systems can affect the electronic structure of the defect as well as the bulk host properties. Fermi level renormalization only accounts for the latter, which is particularly detrimental in the case of shallow defect levels that are well-characterized by delocalized bulk states.

The Projection Shift method is more computationally expensive than other correction methods, though much less so than the actual DFT calculations for a defect. However, by shifting energy levels based on projections onto bulk states, it is possible to roughly estimate the effect of post-DFT physics on transition levels. The limitation of Projection Shifting is that the projection of defect levels onto conduction and valence bands clearly does not provide complete information about how the energy of the defect level will shift when extrapolating from DFT to a more accurate band structure. The chemical environment of an electron in a given state is significantly different in the defect than in the bulk, so the correction to the energy could be very different. One way to account for the local chemical environment around the defect is to partly base the shift in energy of defect levels on the projection onto local atomic wavefunctions around the defect. Local atomic states can have different energies in semi-local and hybrid DFT, which will affect the transition level predictions of localized defect levels.

While the abstract formulations of the methods here have been developed before, pawpyseed is the first implementation of standard tools for performing perturbation-extrapolation corrections for point defect calculations. These tools can potentially improve defect transition level predictions of semi-local DFT point defect calculations enough to be useful in high-throughput screening applications, for which more accurate approaches like hybrid DFT are too computationally expensive.

\section{Methods}

\subsection{Pawpyseed implementation details}

Pawpyseed is written in Python and C. All user interface is written in Python, and computationally expensive tasks are performed by interfacing with a C extension via Cython \cite{behnel2010cython}. The Intel Math Kernel Library is used for linear algebra routines in C, and OpenMP \cite{openmp} is used to implement shared memory parallelization. Pawpyseed reads the output of VASP calculations \cite{vasp}. Pseudo wavefunctions are read from WAVECAR files using a method based on the WaveTrans program \cite{Feenstra}, and data about augmentation regions is read from pseudopotential (POTCAR) files. In addition to the core Python library, pawpyseed relies heavily on NumPy and SciPy \cite{scipy} (for numerical operations), pymatgen \cite{pymatgen} (for storing and manipulating structures, reading VASP input and output files, and performing symmetry analysis), Matplotlib \cite{matplotlib} (for visualization), and SymPy \cite{sympy} (for calculating and storing Gaunt coefficients for overlap operator routines). The code also makes use of the spherical Bessel transform algorithm developed by Talman \cite{TALMAN}.

Because pawpyseed abstracts all high-performance routines to its C extension, complex routines can be performed with single Python calls. The following code snippet reads the pseudo wavefunctions for two VASP calculations (for a bulk structure and defect structure), sets up the projector function and partial wave components, and projects band 0 of the defect structure onto each of the bands of the bulk structure for each k-point and spin. Further examples and documentation are available on the pawpyseed website (Section \ref{code_avail}).

\begin{verbatim}
    from pawpyseed.core.projector import Wavefunction, Projector
    bulk = Wavefunction.from_directory('bulk')
    defect = Wavefunction.from_directory('defect')
    pr = Projector(defect, bulk)
    res = pr.single_band_projection(0)
\end{verbatim}

\subsection{Overlap operators in pawpyseed} \label{overlap_operators}

Pawpyseed calculates the overlaps between PAW wavefunctions
from different structures \cite{blochl}:
\begin{equation}
\ket{\psi_{n\mathbf{k}}}=\ket{\widetilde{\psi}_{n\mathbf{k}}}+
\sum_{i}(\ket{\phi_{i}}-\ket{\widetilde{\phi}_{i}})
\braket{\widetilde{p}_{i}|\widetilde{\psi}_{n\mathbf{k}}}
\label{eq:pawwf}
\end{equation}

There are four options for evaluating these overlaps in pawpyseed. The first, which is used for this project and is the default in pawpyseed, is to individually project the components $\ket{\widetilde{\psi}_{n\mathbf{k}}}$ and $(\ket{\phi_{i}}-\ket{\widetilde{\phi}_{i}})\braket{\widetilde{p}_{i}|\widetilde{\psi}_{n\mathbf{k}}}$ of one wavefunction onto the components of the other. The numerical techniques for this method are complex, and therefore we relegate them to the SI and the technical reference on the pawpyseed website. For this method, aliasing in real space integration is restricted to the overlap between the $\ket{\phi_{i}}-\ket{\widetilde{\phi}_{i}}$ terms and the pseudo wavefunction $\ket{\widetilde{\psi}_{n\mathbf{k}}}$.

The second method is a variation on the first, where the sum $\sum_{i}(\ket{\phi_{i}}-\ket{\widetilde{\phi}_{i}}) \braket{\widetilde{p}_{i}|\widetilde{\psi}_{n\mathbf{k}}}$ is evaluated in real space and then transformed into reciprocal space to project onto $\ket{\widetilde{\psi}_{n\mathbf{k}}}$. Because the $\ket{\phi_{i}}-\ket{\widetilde{\phi}_{i}}$ terms are not multiplied by $\ket{\widetilde{\psi}_{n\mathbf{k}}}$ directly in real space, this method can be used to prevent integration aliasing at the expense of using more memory. We chose not to use this method here because it is still in development, but tests show the choice between these methods does not significantly affect the transition level results.

The third method is to evaluate Equation \ref{eq:pawwf} in real space and then integrate the product of the two wavefunctions over the unit cell. This method requires a denser realspace grid and therefore is more computationally expensive.

The fourth method is to ignore all components except the pseudo wavefunction $\ket{\widetilde{\psi}_{n\mathbf{k}}}$, which is a sum of plane waves and therefore easy to integrate. Because the pseudo wavefunctions are not orthonormal, this method is not recommended and has no guaranteed precision. It is provided as an option in the code because it is very fast and can be used for quick preliminary analysis in some cases.

There are several precision considerations for these integrals, particularly for the first method. For example, any incompleteness in the set of pseudo partial waves $\ket{\widetilde{\phi}_{i}}$ will lead to normalization errors, and the high-frequency components of the all electron partial waves $\ket{\phi_{i}}$ will cause aliasing when integrated in real space for any reasonable grid size. However, these errors tend to be small, and basic tests suggest the precision of the first method above is better than 0.01 for all bands except high conduction bands, which are poorly described by the pseudo partial waves. This precision is adequate for the corrections presented here, and the precision can be improved at the cost of computational efficiency for applications where high precision is necessary.

\subsection{Perturbative band shifting correction implementation} \label{projshift}
For the purpose of benchmarking, a very basic perturbative band shifting method is presented here and referred to as the ``Projection Shift." Because the perturbative band shifting correction adjusts single-particle energy levels, it is nontrivial to decide the correction to the total energy, even for a simplified correction like Equation \ref{eq:pt_corr}. For this work, it is assumed that the shifted single-particle energies $e_{n,\mathbf{k}}$ given by Equation \ref{eq:pt_corr} only apply if $e_{n,\mathbf{k}}$ is in the band gap after being shifted (i.e. it is a defect level). Otherwise, the energy shifts entirely with the set of bands in which it is located (valence or conduction). This is reasonable because Equation \ref{eq:pt_corr} should only be applied to defect levels, but all of the bands shift in energy going from the GGA to hybrid level of theory.

To avoid counting energy shifts from occupied valence bands, a reference energy $N_{VB,X,0}\Delta E_{VBM}$ is chosen, where $N_{VB,X,0}$ is the number of electrons in the valence band of the neutral charge state of the defect. Note that this reference energy does not affect transition levels because transition levels are only dependent on the energy difference between charge states. This method prevents large double counting errors for the Coulomb energy while accounting for the energy shifts of the valence band, conduction band, and defect levels.
\begin{align}
E_{corr,shift} =& -N_{VB,X,0}\Delta E_{VBM}
+ \sum_{n\in BG} \Delta e_{n} \label{eq:projshift}\\
&+ \sum_{n\in VB,\mathbf{k}} \omega_{\mathbf{k}} f_{n,\mathbf{k}} \Delta E_{VBM}
+ \sum_{n\in CB,\mathbf{k}} \omega_{\mathbf{k}} f_{n,\mathbf{k}} \Delta E_{CBM} \notag \\
\Delta e_{n} =& \sum_{\mathbf{k}}
\omega_{\mathbf{k}} f_{n,\mathbf{k}}
(c_{n,\mathbf{k}}\Delta E_{CBM}
+ v_{n,\mathbf{k}}\Delta E_{VBM}) \label{eq:singleshift1}
\end{align}
$\omega_{\mathbf{k}}$ are k-point weights and $f_{n,\mathbf{k}}$ are occupations. BG, VB, and CB are the band gap, valence bands, and conduction bands at the hybrid level of theory, respectively. To include all possible defect levels, each band is categorized by its k-point-averaged energy relative to the \emph{hybrid DFT} band edges and \emph{before} it is shifted by $\Delta e_n$. The averaging is performed because defect levels do not have dispersion in the true physical band structure. Spin polarization is taken into account by replacing the band index $n$ with band and spin indices $n,\sigma$.

An alteration of this method has been implemented that measures how localized a defect level is and then performs a fraction of the energy shift (Equation \ref{eq:pt_corr}) which is larger for more delocalized levels. We define $\chi_n$ as the fraction of the density of state $n$ that is localized on the defect atom and its first set of nearest neighbors. $\chi_n$ can be determined from the site-projected character output from a DFT calculation (PROCAR for VASP). This function is valued between 0 and 1, so it can serve as a fraction of the Projection Shift to perform on each level. This gives an adjusted correction:
\begin{equation}
\Delta e_{n} = (1-\chi_n) \sum_{\mathbf{k}}
\omega_{\mathbf{k}} f_{n,\mathbf{k}}
(c_{n,\mathbf{k}}\Delta E_{CBM}
+ v_{n,\mathbf{k}}\Delta E_{VBM}) \label{eq:singleshift2}
\end{equation}
We refer to this second method as ``Delocalized-State Projection Shifting".

\subsection{DFT calculation details} \label{dftcalc}
All DFT calculations were performed using the VASP code \cite{vasp,vasp1,vasp2,vasp3} with the Perdew-Burke-Ernzerhof (PBE) generalized gradient approximation (GGA) functional \cite{ggapbe1,ggapbe2} and the PAW method \cite{vasp4,blochl}. For hybrid functional bulk calculations, the Heyd-Scuseria-Ernzerhof (HSE) method \cite{Heyd2003} was used with the standard HSE06 \cite{Krukau2006} tuning parameters. All k-point grids used were $\Gamma$-centered to effectively sample band extrema, and all calculations were performed with symmetry off to allow the environment of the defect to relax out of its initial symmetry.

The boron, phosphorus, copper, and sulfur substitutionals and single vacancy were embedded in a 216-atom supercell of silicon using PyCDT \cite{pycdt}. PyCDT was also used to generate charge states for analysis. Energy convergence tests were performed with the number of k-points. For the total cohesive energy of the bulk crystal, a difference of 0.02 eV ($1\times 10^{-4}$ eV/atom) was found between a 2x2x2 and 3x3x3 k-point mesh for the bulk supercell, so a 2x2x2 k-point mesh was used for bulk and defect supercell calculations. The energy cutoff was set to 520 eV. An ionic relaxation and total energy calculation were performed for each defect and charge state. The static dielectric constant was also calculated for the bulk silicon crystal using VASP and found to be 12.52. For the silicon vacancy, even with symmetry turned off, the structure does not relax to a stable state because it is caught in a metastable minimum. The final symmetry is different for each charge state, but as an example, the neutral defect has a tetragonally distorted structure with $D_{2d}$ symmetry \cite{Wright2006}. To attain the correct structure, the positions of the four atoms nearest the vacancy site were perturbed manually to break the structure from the metastable minimum and break symmetry before the calculation was run.

To calculate DFT and hybrid band edges, a DFT calculation was run on a single 2-atom silicon primitive cell with a 7x7x7 k-point mesh, once with the GGA functional and once with the HSE06 functional. The GGA band gap calculated in this study is 0.61 eV, in good agreement with the Materials Project GGA band gap of 0.612 eV \cite{materialsproject}. The HSE06 band gap is 1.19 eV, in good agreement with the experimental 0 K value of 1.17 eV \cite{Bludau1974}. For our calculations, $\Delta E_{VBM}=-0.36$ eV and $\Delta E_{CBM}=0.22$ eV.

\section{Code Availability} \label{code_avail}
Analysis scripts will be published on Github and linked to in a future manuscript draft. The pawpyseed source code and installation instructions can be found at \url{https://github.com/kylebystrom/pawpyseed}. The software documentation, along with a technical reference guide and run-time scaling analysis for computationally expensive routines, can be found at \url{https://kylebystrom.github.io/pawpyseed/}.

\section{Data Availability} \label{data_avail}
Electronic structure data will be published to the NOMAD database in the near future, and this manuscript will be updated with access information.

\section{Acknowledgements}

This work was funded by the U.S. Department of Energy, Office of Science, Office of Basic Energy Sciences, Materials Sciences and Engineering Division under Contract No. DE-AC02-05-CH11231 (Materials Project program KC23MP). This research used the Savio computational cluster resource provided by the Berkeley Research Computing program at the University of California, Berkeley (supported by the UC Berkeley Chancellor, Vice Chancellor for Research, and Chief Information Officer).

\section{Author Contributions}

KB wrote pawpyseed and data analysis scripts, performed the DFT calculations, and wrote the manuscript. DB developed the idea for the project, helped refine the correction methods, and assisted with interpretation of results. MA supervised the code development and initial collection of results. SD and KP supervised code and method refinement and presentation of results. All authors reviewed the manuscript.

\printbibliography

\end{document}


\maketitle

\section{Overlap Operator Formalism}

The operator that maps a pseudo (PS) wavefunction to an
all electron (AE) wavefunction in the PAW method is as follows \cite{blochl}:
\begin{equation}
T=1+\sum_a\sum_l\sum_m\sum_{\epsilon}(\ket{\phi_{a\ind}}-\ket{\widetilde{\phi}_{a\ind}})
\bra{\widetilde{p}_{a\ind}}=1+\sum_i(\ket{\phi_i}-\ket{\widetilde{\phi}_i})
\bra{\widetilde{p}_i}
\label{eq:T}
\end{equation}
The following definitions and conditions are used:
\begin{itemize}
    \item $\phi_{a\ind}$ are AE partial waves.
    \item $\widetilde{\phi}_{a\ind}$ are PS partial waves. \item $\widetilde{p}_{a\ind}$ are projector functions.
    \item $a$ are the site indices of each atom in the structure.
    \item $l$, $m$, and $\epsilon$ specify a spherical harmonic and energy quantum number which uniquely specify a partial wave at a given atomic site.
    \item Projector functions $\widetilde{p}_{a\ind}$ are localized within a cutoff radius $r_c$ around the nucleus of site $a$.
    \item $\widetilde{\phi}_{a\ind}=\phi_{a\ind}$ outside an augmentation radius $r_a$.
\end{itemize}

To evaluate operators in the PAW formalism, one defines a pseudo operator $\widetilde{A}$ for each operator $A$ such that $\bra{\psi}A\ket{\psi}=\bra{\widetilde{\psi}}\widetilde{A}\ket{\widetilde{\psi}}$. Because $\ket{\psi}=T\ket{\widetilde{\psi}}$, one can write \cite{blochl}:
\begin{equation}
\widetilde{A}=T^{\dagger}AT
\label{eq:op}
\end{equation}
One can then plug Equation \ref{eq:T} into Equation \ref{eq:op} to find
\begin{align}
\widetilde{A} = & [1+\sum_i\ket{\widetilde{p}_i}(\bra{\phi_i}-\bra{\widetilde{\phi}_i})]
A[1+\sum_j(\ket{\phi_j}-\ket{\widetilde{\phi}_j})
\bra{\widetilde{p}_j}] \label{eq:res1} \\
\widetilde{A} = & A+\sum_i\ket{\widetilde{p}_i}(\bra{\phi_i}-\bra{\widetilde{\phi}_i})A
+\sum_j A(\ket{\phi_j}-\ket{\widetilde{\phi}_j})\bra{\widetilde{p}_j} \label{eq:res} \\ \notag
& +\sum_i\sum_j\ket{\widetilde{p}_i}(\bra{\phi_i}-\bra{\widetilde{\phi}_i})A
(\ket{\phi_j}-\ket{\widetilde{\phi}_j})\bra{\widetilde{p}_j}
\end{align}
When the operator $A$ is local, then $\sum_j\ket{\widetilde{\phi}_j}\bra{\widetilde{p}_j}=1$, which reduces Equation \ref{eq:res} to a simpler form for local operators \cite{blochl}:
\begin{equation}
\widetilde{A}=A+\sum_i\sum_j\ket{\widetilde{p}_i}(\bra{\phi_i}A\ket{\phi_j}
-\bra{\widetilde{\phi_i}}A\ket{\widetilde{\phi_j}})\bra{\widetilde{p}_j}
\label{eq:loc1}
\end{equation}

The following section introduces an equation for the overlap operator between one Kohn-Sham single particle state of one structure $R$ and one Kohn-Sham single particle state of another structure $S$, where $R$ and $S$ share a common lattice and the DFT PS wavefunctions are constructed with the same plane-wave basis set in the PAW formalism. The goal is to present this overlap operator in a form convenient for computation.

Starting with Equation \ref{eq:res} for any single-particle operator in the PAW formalism and replacing $A$ with unity gives the overlap operator (note that the summation over $i$ is for structure $R$, and the summation over $j$ is for structure $S$):
\begin{equation}
\begin{split}
\braket{\psi_{Rn_1\mathbf{k}}|\psi_{Sn_2\mathbf{k}}} & =O_0+O_1+O_2+O_3\\
O_0 & =\braket{\widetilde{\psi}_{Rn_1\mathbf{k}}|\widetilde{\psi}_{Sn_2\mathbf{k}}}\\
O_1 & =\sum_i\braket{\widetilde{\psi}_{Rn_1\mathbf{k}}|\widetilde{p}_i}(\bra{\phi_i}-\bra{\widetilde{\phi}_i})
\ket{\widetilde{\psi}_{Sn_2\mathbf{k}}}\\
O_2 & =\sum_j\bra{\widetilde{\psi}_{Rn_1\mathbf{k}}}(\ket{\phi_j}-\ket{\widetilde{\phi}_j})
\braket{\widetilde{p}_j|\widetilde{\psi}_{Sn_2\mathbf{k}}}\\
O_3 & =\sum_i\sum_j\braket{\widetilde{\psi}_{Rn_1\mathbf{k}}|
\widetilde{p}_i}(\bra{\phi_i}-\bra{\widetilde{\phi}_i})
(\ket{\phi_j}-\ket{\widetilde{\phi}_j})\braket{\widetilde{p}_j|\widetilde{\psi}_{Sn_2\mathbf{k}}}
\end{split}
\label{eq:ov1a}
\end{equation}

It is important to simplify the calculation of the other terms in equation \ref{eq:ov1a} as much as possible because the calculation can be computationally expensive, and the number of necessary calculations for projecting onto an entire basis set scales with the number of sites times the size of the basis set. One major simplification is that if a site $a$ in structure $R$ and site $b$ in structure $S$ have the same species and position, $a$ and $b$ will only have overlapping augmentation regions with each other and no other sites. Then, the terms in $O_1,O_2,O_3$ on site $a$ or $b$ (denoted $O_{1a},O_{2a},O_{3a}$) simplify to the local operator solution derived by Blochl:
\begin{equation}
O_{1a}+O_{2a}+O_{3a}=\sum_{l,m}\sum_{\epsilon_1}\sum_{\epsilon_2}
\braket{\widetilde{\psi}_{Rn_1\mathbf{k}}|\widetilde{p}_{a\ind _1}}
(\braket{\phi_{a\ind _1}|{\phi_{a\ind _2}}}
-\braket{\widetilde{\phi}_{a\ind _1}|\widetilde{\phi}_{a\ind _2}})
\braket{\widetilde{p}_{a\ind _2}|\widetilde{\psi}_{Sn_2\mathbf{k}}}
\end{equation}
All three terms must be evaluated in full for the other sites, but terms in $O_3$ where $i$ and $j$ correspond to sites with non-overlapping augmentation spheres vanish. Therefore,
\begin{align}
\braket{\psi_{Rn_1\mathbf{k}}|\psi_{Sn_2\mathbf{k}}}&=O_0+O_M+O_R+O_S+O_N
\label{eq:overlap}\\
O_M&=\sum_{i,j\in M_{RS}}
\braket{\widetilde{\psi}_{Rn_1\mathbf{k}}|\widetilde{p}_{i}}
(\braket{\phi_{i}|{\phi_{j}}}
-\braket{\widetilde{\phi}_{i}|\widetilde{\phi}_{j}})
\braket{\widetilde{p}_{j}|\widetilde{\psi}_{Sn_2\mathbf{k}}}\\
O_R&=\sum_{i\in N_R}\braket{\widetilde{\psi}_{Rn_1\mathbf{k}}|\widetilde{p}_i}
(\bra{\phi_i}-\bra{\widetilde{\phi}_i})
\ket{\widetilde{\psi}_{Sn_2\mathbf{k}}}\\
O_S&=\sum_{j\in N_S}\bra{\widetilde{\psi}_{Rn_1\mathbf{k}}}(\ket{\phi_j}
-\ket{\widetilde{\phi}_j})
\braket{\widetilde{p}_j|\widetilde{\psi}_{Sn_2\mathbf{k}}}\\
O_N&=\sum_{i,j\in N_{RS}}\braket{\widetilde{\psi}_{Rn_1\mathbf{k}}|
\widetilde{p}_i}(\bra{\phi_i}-\bra{\widetilde{\phi}_i})
(\ket{\phi_j}-\ket{\widetilde{\phi}_j})
\braket{\widetilde{p}_j|\widetilde{\psi}_{Sn_2\mathbf{k}}}
\end{align}
with the following definitions:
\begin{itemize}
    \item  $M_{RS}$ is the set of pairs of identical sites in $R$ and $S$.
    \item $N_R$ and $N_S$ are the sets of sites in $R$ and $S$ not in $M_{RS}$.
    \item $N_{RS}$ is the set of pairs of sites not in $M_{RS}$ with overlapping augmentation regions
\end{itemize}

\section{``Desymmetrization" Routines}

Since changing the basis of a lattice
(atoms and atomic positions) can change the space group, and
because DFT calculations reduce the k-point sampling space based
on symmetry operations, it is important for a code which calculates
overlap operators of wavefunctions from different structures to
be able to derive a wavefunction at one k-point from a wavefunction
at a symmetrically identical k-point. Two k-points $\mathbf{k}$
and $\mathbf{k'}$ are symmetrically identical if 
$\mathbf{k'}=\Theta \mathbf{k}$, where $\Theta$ is the
non-translation component of 
a space group operation $R=T\Theta$ of the crystal,
where $T$ is the translation. For nonmagnetic systems,
they are also symmetrically identical if related
by time inversion ($\tau: t \rightarrow -t$).

Because $R$ commutes with $H$, the above condition guarantees that
\begin{align}
H\psi_{n\mathbf{k}} &= E_{n\mathbf{k}}\psi_{n\mathbf{k}}
\label{eq:known_eig} \\
HR\psi_{n\mathbf{k}} &= E_{n\mathbf{k}}R\psi_{n\mathbf{k}}
\label{eq:unknown_eig}
\end{align}
Since the k-point of $R\psi_{n\mathbf{k}}$ is $\mathbf{k'}$,
the eigenfunctions at $\mathbf{k'}$ can be specified as:
\begin{equation}
\psi_{n\mathbf{k'}}=R\psi_{n\mathbf{k}}
\label{eq:ansatz}
\end{equation}
Next, the wavefunctions are expressed as a sum of plane waves:
\begin{equation}
\psi_{n\mathbf{k}}(\mathbf{r})=\sqrt{\frac{1}{V}}
e^{i\mathbf{k}\cdot \mathbf{r}}\sum_{\mathbf{G}} C_{n,\mathbf{k},\mathbf{G}}
e^{i\mathbf{G}\cdot \mathbf{r}}
\label{eq:pseudowf}
\end{equation}
Plugging Equation \ref{eq:pseudowf} into Equation \ref{eq:ansatz}
and taking $T=\Delta r$ to be the translational component of $R$,
a condition on the plane-wave constants can be derived as shown:
\begin{align}
\psi_{n\mathbf{k}}(\mathbf{r}) &=
R\psi_{n\mathbf{k}}(R\mathbf{r}) \notag \\
\psi_{n\mathbf{k}}(\mathbf{r}) &=
\psi_{n\mathbf{k'}}\rp \notag \\
e^{i\mathbf{k}\cdot \mathbf{r}}\sum_{\mathbf{G}}
C_{n,\mathbf{k},\mathbf{G}} e^{i\mathbf{G}\cdot \mathbf{r}}
&= e^{i\mathbf{\Theta k}\cdot \rp}\sum_{\mathbf{G}}
C_{n,\Theta \mathbf{k},\mathbf{G}} e^{i\mathbf{G}\cdot \rp} \notag \\
\sum_{\mathbf{G}} C_{n,\mathbf{k},\mathbf{G}} e^{i\mathbf{G}\cdot \mathbf{r}}
&= e^{i\Theta \mathbf{k}\cdot \Delta \mathbf{r}}\sum_{\mathbf{G}}
C_{n,\Theta \mathbf{k},\mathbf{G}} e^{i\mathbf{G}\cdot \rp} \notag \\
\sum_{\mathbf{G}} C_{n,\mathbf{k},\mathbf{G}} e^{i\mathbf{G}\cdot \mathbf{r}}
&= e^{i\Theta \mathbf{k}\cdot \Delta \mathbf{r}}\sum_{\mathbf{G}}
C_{n,\Theta \mathbf{k},\Theta \mathbf{G}} e^{i\Theta \mathbf{G}\cdot \rp} \notag \\
\sum_{\mathbf{G}} C_{n,\mathbf{k},\mathbf{G}} e^{i\mathbf{G}\cdot \mathbf{r}}
&= \sum_{\mathbf{G}}
C_{n,\Theta \mathbf{k},\Theta \mathbf{G}} e^{i\mathbf{G}\cdot \mathbf{r}}
e^{i\Theta \mathbf{(k+G)}\cdot \Delta \mathbf{r}} \notag \\
C_{n,\mathbf{k},\mathbf{G}}
&= e^{i\Theta \mathbf{(k+G)}\cdot \Delta \mathbf{r}}
C_{n,\Theta \mathbf{k},\Theta \mathbf{G}}
\label{eq:symmetry_condition}
\end{align}

This simple relationship allows a PS wavefunction at $\mathbf{k}$
to be quickly mapped to a PS wavefunction at $\Theta \mathbf{k}$.
Similarly, if time reversal symmetry holds:
\begin{equation}
C_{n,\mathbf{k},\mathbf{G}} = C^{*}_{n,-\mathbf{k},-\mathbf{G}}
\end{equation}
The complex conjugation occurs because time reversal is an
antilinear operator.

\section{Overlap Operator Implementation Details}

This section presents the numerical methods
used to calculate overlap operators in the form of Equation \ref{eq:overlap}.
Table \ref{tab:runtime} gives detailed run-time scaling for each portion
of the routine.

\begin{table}[t]
\center
\caption{Runtime scaling functions for each component of the code and
definitions for shorthand symbols to express runtime. Approximate
scales with the number of electrons are also shown.}
\label{tab:runtime}

\begin{tabular}{|c|c|c|}
\hline
Computational Task & $\Theta$ & Frequency*\\
\hline
$O_0$ & $BKSW \sim n^2$ & per band\\
$O_M$ & $BKSNP \sim n^2$ & per band\\
$O_R$ and $O_S$ & $BKSNP \,\,\mathrm{\textbf{or}}\,\, BKSW \sim n^2$ & per band\\
$O_N$ & $BKSNP \sim n^2$ & per band\\
$\braket{\widetilde{p}_i|\widetilde{\psi}_{n\mathbf{k}}}$ & $BKSF\mathrm{log}(F)
\sim n^2\mathrm{log}(n)$ & per structure\\
$(\bra{\phi_i}-\bra{\widetilde{\phi}_i})
\ket{\widetilde{\psi}_{n\mathbf{k}}}$ & $BKSF\mathrm{log}(F)
\sim n^2\mathrm{log}(n)$ & per structure\\
spherical Bessel transform partial waves & $EPG\mathrm{log}(G) \sim 1$ & per structure\\
projections for overlapping aug. spheres & $NP^2G\mathrm{log}(G) \sim n$ & per structure pair\\
\hline
\end{tabular}

\begin{tabular}{c|c}
Symbol & Definition\\
\hline
B & number of bands\\
E & number of elements\\
F & size of FFT grid\\
G & size of partial wave radial grid\\
K & number of k-points\\
N & number of sites**\\
P & number of projector functions\\
S & number of spin states\\
W & number of plane waves\\
n & number of electrons (approximate scaling)
\end{tabular}\\
\justify *The frequency refers to how often the routine is called. ``Per band"
indicates that the routine runs once every time a band from one structure
is projected onto all the bands of a basis structure. ``Per structure"
indicates a setup routine used to set up the wavefunctions for a structure.
``Per structure pair" is a setup routine run once for each pair of structures
for which the overlap operators are to be calculated.\\
**Number of sites flexibly refers to the number of sites relevant to the calculation,
which worst-case scales with the total number of sites in the structure. For example,
calculating $O_M$ and $O_N$ only require the sites in sets $M_{RS}$ and $N_{RS}$, respectively.
\end{table}

\subsection{Overlap of Pseudo Wavefunctions ($O_0$)}

The PS wavefunction is a summation of plane waves (Equation \ref{eq:pseudowf}), so the overlap between two PS wavefunctions per unit cell can be written as
\begin{equation}
O_0=\braket{\widetilde{\psi}_{n_1\mathbf{k}_1}|\widetilde{\psi}_{n_2\mathbf{k}_2}}
=\delta_{\mathbf{k}_1,\mathbf{k}_2}
\sum_{\mathbf{G}} C^*_{n_1\mathbf{k}_1\mathbf{G}}C_{n_2\mathbf{k}_2\mathbf{G}}
\end{equation}

\subsection{Concentric Augmentation Spheres ($O_M$)}

Integrals of the type $\braket{\widetilde{\psi}_{n\mathbf{k}}|\widetilde{p}_{i}}$ between a PS wavefunction and projector function are evaluated by projecting $\widetilde{\psi}$ onto a real-space FFT grid and evaluating the projector functions $\widetilde{p}_{i}$ at the points on the grid. This is a typical real-space projector function evaluation routine like that used in VASP with LREAL=TRUE \cite{vasp,vasp1,vasp2,vasp3,vasp4}. Integrals of the type $\braket{\phi_{i}|{\phi_{j}}}$ between partial waves are evaluated by simple radial integration. This is possible because the augmentation spheres are concentric, so the spherical harmonics for the partial waves are orthonormal.

\subsection{Partial Waves Overlapping with Pseudo Wavefunction ($O_R$, $O_S$)}

These integrals require projecting a smoothly varying PS wavefunction onto a rapidly varying AE partial wave. There are two options for doing this in pawpyseed. For both methods, frequency components higher than the cutoff frequency of the FFT grid are removed to avoid aliasing. This preserves the accuracy of the integral because the PS wavefunction has a frequency cutoff smaller than the cutoff of the FFT grid, and plane waves of different frequencies are orthogonal.

The first method is to simply follow the method for real-space projector evaluation with the partial wave differences $(\bra{\phi_i}-\bra{\widetilde{\phi}_i})$ in place of the projector functions. For a grid with the same density as used for projector evaluation, the integral may contain some aliasing because the maximum frequency of the product of the filtered partial waves and PS wavefunction is the sum of the maximum frequencies in the partial waves and PS wavefunctions. Theoretically, one could filter all frequency components down to $G_{cut}$ of the DFT calculation, but in practice we find this reduces the precision of the integral, possibly because the partial waves are not accurately represented in real space by such a low cutoff. Therefore, this method theoretically has some aliasing, but we have not observed it being significant for $PREC=Accurate$ VASP calculations. This method scales as $\Theta (BKSNP \sim n^2)$ (see Table \ref{tab:runtime}).

The second method is to evaluate the sum over partial wave differences $\sum_{i}(\ket{\phi_{i}}-\ket{\widetilde{\phi}_{i}}) \braket{\widetilde{p}_{i}|\widetilde{\psi}_{n\mathbf{k}}}$ in real space for each PS wavefunction and then Fourier transform it into reciprocal space. These terms can then be projected onto PS wavefunctions from the other structure in reciprocal space. This method has the same runtime scaling as above and theoretically avoids aliasing at the expense of increased memory usage, although in practice we have not observed a significant difference from the first method. This method scales as $\Theta (BKSW \sim n^2)$ (see Table \ref{tab:runtime}).

Note: To filter high-frequency components from the plane waves, the $O(NlogN)$ NUMSBT algorithm developed by Talman \cite{TALMAN} is used. Then, all frequency components greater than the 1-dimensional FFT grid density are set to 0, and the ``filtered" partial waves can be transformed back into real space, also using the NUMSBT algorithm. This results in smooth partial waves for which $(\bra{\phi_i}-\bra{\widetilde{\phi}_i})\ket{\widetilde{\psi}_{n\mathbf{k}}}$ can be evaluated in real space.

\subsection{Partial Wave Overlap on Non-Orthogonal Augmentation Spheres ($O_N$)}

The $O_N$ term appears similar to the $O_M$ term, except that the integrals $(\bra{\phi_i}-\bra{\widetilde{\phi}_i})(\ket{\phi_j}-\ket{\widetilde{\phi}_j})$ contain partial waves centered at different sites. However, transforming these partial waves into reciprocal space using the spherical Bessel transform allows the overlap integrals to be evaluated using Equation 47 in Talman's NUMSBT paper \cite{TALMAN}. Evaluating this equation requires the Gaunt coefficients, which are calculated and stored using SymPy \cite{sympy}.

\section{Transition levels for correction schemes without band filling}

See Table \ref{tab:tlcomp}.

\begin{table}[t]
\caption{Errors of predicted transition levels in eV for the DFT calculations. Format: ``level in eV (error as percentage of band gap)." The band gap for the Fr method is 0.61 eV, while for the other methods it is 1.19 eV. The HSE06 band gap in other studies is 1.16-1.19 eV. For this set of correction schemes, \emph{band filling corrections were not used}.}
\begin{tabular}{|c|c|c|c|c|c|c|}
\hline
Transition Level & Reference & Fr & Fr+FLR & Fr+FLR+PrS & Fr+FLR+DPrS\\
\hline
Vac ++/+	& 0.14 \cite{Spiewak2013}	& -0.13 (-33\%)	& 0.24 (8\%)	& -0.05 (-16\%)	& 0.09 (-4\%) \\
Vac +/0		& 0.02 \cite{Spiewak2013}	& -0.00 (-2\%)	& 0.36 (28\%)	& -0.02 (-3\%)	& -0.01 (-2\%) \\
P +/0		& 1.07 \cite{Jagannath1981}	& 0.49 (-15\%)	& 0.86 (-24\%)	& 1.07 (-6\%)	& 1.04 (-8\%) \\
B 0/-		& 0.04 \cite{Fischer1983}	& 0.08 (9\%)	& 0.44 (33\%)	& 0.08 (3\%)	& 0.10 (4\%) \\
Cu +/0		& 0.20 \cite{Sharan2017}	& 0.05 (-9\%)	& 0.41 (18\%)	& 0.13 (-6\%)	& 0.32 (10\%) \\
Cu 0/-		& 0.54 \cite{Sharan2017}	& 0.32 (7\%)	& 0.68 (11\%)	& 0.41 (-12\%)	& 0.58 (3\%) \\
Cu -/--		& 0.97 \cite{Sharan2017}	& 0.63 (20\%)	& 0.99 (0\%)	& 0.77 (-19\%)	& 0.90 (-8\%) \\
S ++/+		& 0.58 \cite{Deak2010}		& 0.10 (-33\%)	& 0.46 (-11\%)	& 0.62 (3\%)	& 0.53 (-5\%) \\
S +/0		& 0.86 \cite{Deak2010}		& 0.43 (-4\%)	& 0.79 (-7\%)	& 0.97 (8\%)	& 0.89 (1\%) \\
\hline
\end{tabular}\\
\label{tab:tlcomp}
\end{table}

\printbibliography